\documentclass[sigconf]{acmart}

\usepackage{draftwatermark}

\SetWatermarkText{Submitted to ACM CCS 2026}
\SetWatermarkScale{0.3}
\SetWatermarkColor[gray]{0.85} 
\AtBeginDocument{%
  }

\usepackage{tikz}
\usetikzlibrary{arrows.meta, positioning}
\setcopyright{acmlicensed}
\copyrightyear{2018}
\acmYear{2018}
\acmDOI{XXXXXXX.XXXXXXX}
\acmConference[Conference acronym 'XX]{Make sure to enter the correct
  conference title from your rights confirmation email}{June 03--05,
  2018}{Woodstock, NY}
\acmISBN{978-1-4503-XXXX-X/2018/06}

\begin{document}

\title{Anonymous YARA Rules Are Not Anonymous}


\author{Usman Rabiu ISAH}
\affiliation{%
  \institution{INSA Centre Val de Loire}
  \city{Bourges}
  \country{France}}
  \affiliation{%
  \institution{Sule Lamido University }
  \city{Kafin Hausa}
  \country{Nigeria}}
\email{usman.isah@insa-cvl.fr}
\orcid{0000-0002-1984-0208}

\author{Laurent BOBELIN}
\orcid{0000-0002-3268-4203}
\affiliation{%
  \institution{INSA Centre Val de Loire}
  \city{Bourges}
  \country{France}}
\email{laurent.bobelin@insa-cvl.fr}

\author{Pascal BERTHOMÉ}
\orcid{0000-0003-4145-6349}
\affiliation{%
  \institution{INSA Centre Val de Loire}
  \city{Bourges}
  \country{France}}
\email{pascal.berthome@insa-cvl.fr}
\renewcommand{\shortauthors}{Anonymous Authors}


\begin{abstract}
 YARA rules are widely shared across threat intelligence communities to enable collective defence against malware.
This practice implicitly assumes that removing metadata (e.g., author fields) sufficiently protects the identity of contributing organisations.
To assess the validity of this assumption, we systematically evaluate how much can be inferred from YARA rule text alone.
Specifically, using a corpus of 23,305 rules from three major public repositories, we train independent classifiers along four stylometric fingerprint dimensions: individual author, source repository, malware family, and temporal drift, using three complementary methods: lexical n-grams (Burrows' Delta), syntactic AST features (Caliskan-Islam), and fine-tuned CodeBERT.
Our results demonstrate that repository origin is almost perfectly recoverable (up to 99\% accuracy), individual authors can be re-identified well above chance (76\%), and malware family classification reaches 95\%.
Comparing the same repository attribution task across full-history and time-restricted subsets reveals a 9–18\% accuracy gap, providing preliminary evidence of temporal drift in repository fingerprints.
To further disentangle content from style, we conduct per-malware family author attribution experiments. Even when the malware family is the same for all samples considered, authors can still be re-identified for five of seven tested families (mean accuracy 74.6\%).

These findings constitute the first systematic demonstration that YARA rule sharing is a measurable OPSEC attack surface, and that metadata removal alone does not mitigate it.
\end{abstract}

\ccsdesc[500]{Security and privacy~Pseudonymity, anonymity and untraceability}
\ccsdesc[300]{Security and privacy~Privacy protections}
\ccsdesc[300]{Security and privacy~Social aspects of security and privacy}
\ccsdesc[100]{Security and privacy~Intrusion/anomaly detection and malware mitigation}
\ccsdesc[100]{Computing methodologies~Supervised learning by classification}

\ccsdesc[500]{Security and privacy~Intrusion detection systems}
\ccsdesc[300]{Security and privacy~Malware and its mitigation}
\ccsdesc[300]{Computing methodologies~Machine learning}
\keywords{YARA rules, authorship attribution, re-identification, stylometric fingerprinting, threat intelligence sharing, CodeBERT}



\maketitle

\section{Introduction}

\label{sec:introduction}

Signature-based intrusion detection systems remain a cornerstone of operational cybersecurity.
YARA is among the most widely used signature languages. It is a pattern-matching language for malware identification and classification ~\cite{alvarez2004yara}.
YARA rules are routinely exchanged through public repositories, Information Sharing and Analysis Centres (ISACs), and inter-organisational partnerships to strengthen collective defence.

This sharing is no longer just a best practice, it is now a regulatory expectation.
The European Union's NIS2 Directive (Directive 2022/2555) entered into force in October 2024. It requires cooperation, information sharing, and coordinated incident response. This mandate applies to operators of essential services in 18 critical sectors~\cite{nis2directive}.
At the European level, sector-specific ISACs help exchange threat intelligence, including detection signatures, among their members. These include the European Energy ISAC (EE-ISAC), the Financial Institutes ISAC (FI-ISAC), the European Rail ISAC (ER-ISAC), and the Maritime ISAC~\cite{enisa2017isac}.
National agencies also publish detection rules in threat advisories, and serve as a trusted intermediary in an ISAC: if one of the ISAC member identify a new threat, the trusted intermediary is in charge of anonymizing the detection content (IoC, detection rules) before sharing it to other partners, because those detection content is sensitive data.

On the technical side, tools like OpenCTI, a platform co-developed by ANSSI, CERT-EU, and the Luatix non-profit, provides infrastructure for storing and disseminating YARA, Sigma, and STIX indicators across organisational boundaries~\cite{opencti}.

In these sharing mechanisms, a common assumption prevails: that removing explicit metadata fields (e.g., the \texttt{author} tag, organisational names) from a rule before publication is sufficient to protect the contributor's identity.
The present work directly challenges this widespread assumption, setting the stage for our core analysis.
We show that YARA rules carry persistent stylistic and structural fingerprints that survive metadata stripping.
These fingerprints can be exploited by machine learning classifiers. An adversary can then recover the source repository of a rule with near-perfect accuracy. In many cases, the individual author can also be identified at rates well above chance.

\textit{Threat model.}
We consider a \emph{passive} adversary who collects anonymised YARA rules from public repositories or ISAC feeds. The threat is particularly acute for organisations that contribute rules through ISACs or government-coordinated sharing programmes, where \emph{contributor anonymity} and \emph{unlinkability} of a producer's rules are explicit design goals of the sharing framework (cf.~NIST SP~800-150's principle of limited attribution~\cite{NIST.SP.800-150}).

\paragraph{Contributions.}
This paper makes the following contributions:

\begin{enumerate}
\item We provide the first study of \textbf{multi-dimensional stylometric fingerprinting} of YARA rules, identifying candidates dimensions and determining if those dimension are attributable. Our empirical results shows YARA rules can carry multiple independent fingerprint dimensions: author, repository, and malware family. We compare different standard  classification methods to identify those fingerprint dimensions: lexical n-grams, syntactic AST features, and fine-tuned CodeBERT. To determine if a dimension is part of the fingerprint, we use a confound-controlled author attribution experiment in which malware family and then both family and repository are held constant. 

\item We also found \textbf{preliminary evidence of temporal drift in repository fingerprints}: on the one repository in our corpus with reliable per-rule timestamps (Neo23x0/signature-base), attribution accuracy on the 2022–2025 subset reaches 99\%, compared with 90\% on the full history for our strongest method (CodeBERT), and drops to 81–83\% for the hand-crafted baselines. We frame this as a suggestive finding that motivates further study on corpora with richer temporal annotation, rather than as a definitive characterisation of convergence across the YARA ecosystem.


\item We provide a reproducible experimental pipeline and dataset methodology for future work in detection rule anonymity.
\end{enumerate}

To the best of our knowledge, this is the first systematic study of authorship and provenance attribution on YARA rules, and the first to demonstrate an externally verifiable institutional attribution chain in the detection-rule ecosystem.

The remainder of this paper is organised as follows.
Section~\ref{sec:rule-production} describes the YARA rule production pipeline and the fingerprints dimensions it introduces. Section~\ref{sec:related-work} reviews related work in source-code authorship attribution and threat-intelligence sharing. Section \ref{sec:problem} formalize the problem we study. Section~\ref{sec:methodology} describes the dataset, features, and classification methods. Section~\ref{sec:results} presents results along the four stylometric fingerprint dimensions we identified. Section~\ref{sec:discussion} discusses implications, and Section~\ref{sec:conclusion} concludes.

\section{How YARA Rules Are Produced}
\label{sec:rule-production}

\begin{figure*}[t]
  \centering
  \begin{tikzpicture}[
      node distance=10mm and 10mm,
      box/.style={draw, rounded corners, align=center,
                  inner sep=6pt, minimum height=9mm},
      small/.style={font=\small},
      arrow/.style={-Latex, thick},
    ]
    \node[box, small] (author) {Authoring\\\& Local Conventions};
    \node[box, small, right=of author] (tool)  {Tooling\\\& Templates};
    \node[box, small, right=of tool]   (norm)  {Normalization\\\& Quality Gates};
    \node[box, small, right=of norm]   (trans) {Translation\\\& Backend Adaptation};
    \node[box, small, right=of trans]  (share) {Distribution\\\& Sharing};

    \draw[arrow] (author) -- (tool);
    \draw[arrow] (tool)   -- (norm);
    \draw[arrow] (norm)   -- (trans);
    \draw[arrow] (trans)  -- (share);

    \node[small, align=center, below=7mm of author] (a1)
          {Stylistic preferences\\(granularity, naming)};
    \node[small, align=center, below=7mm of tool]   (a2)
          {Tool fingerprints\\(boilerplate, ordering)};
    \node[small, align=center, below=7mm of norm]   (a3)
          {Standardized structure\\(consistent patterns)};
    \node[small, align=center, below=7mm of trans]  (a4)
          {Systematic rewrites\\(mappings, expansions)};
    \node[small, align=center, below=7mm of share]  (a5)
          {Observable outputs\\(shared artifacts)};

    \draw[arrow] (author) -- (a1);
    \draw[arrow] (tool)   -- (a2);
    \draw[arrow] (norm)   -- (a3);
    \draw[arrow] (trans)  -- (a4);
    \draw[arrow] (share)  -- (a5);

    \node[small, above=8mm of share] (obs)
          {\textbf{Attacker observes shared rules}};
    \draw[arrow] (obs) -- (share);
  \end{tikzpicture}
  \caption{Generic rule transformation and deployment pipeline. }
  \Description{Each stage introduces persistent structural or stylistic fingerprints that
  remain observable after metadata removal.}
  \label{fig:rule-transformation-pipeline}
\end{figure*}
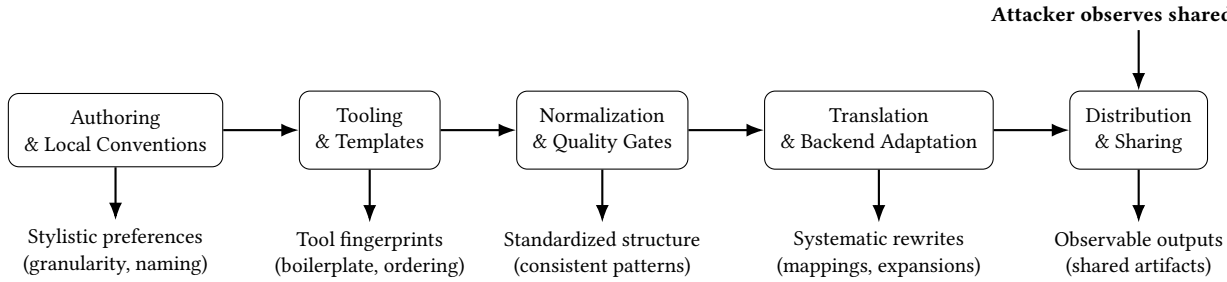

To understand why shared YARA rules can leak the identity of their producer, we first describe how such rules come into existence. Detection rules are not static, hand-crafted artefacts. In practice they are the product of a multi-stage lifecycle that begins with the observation of malicious activity and ends with deployment and maintenance in operational environments~\cite{yarGen}. Along the way, a combination of human analysts, automated generators, organisational policies, and format-conversion tools transform the rule repeatedly
before it is shared. Figure~\ref{fig:rule-transformation-pipeline} summarises this generic pipeline. The remainder of this section walks through each of its five stages and explains why each one can embed persistent structural or stylistic artifacts that survive metadata removal.

\paragraph{Stage 1: Authoring and local conventions.}
A YARA rule begins life when an analyst encodes the
distinguishing features of a malicious artefact as a signature.
Even when targeting the same threat, two analysts produce
substantively different rules: they select different strings,
choose different variable naming schemes (\texttt{\$s1},
\texttt{\$str\_payload}, \texttt{\$marker\_01}), decompose boolean
logic differently, and prioritise different indicators. These
choices reflect each analyst's training and preferences,
producing stable conventions across their corpus that aggregate
into an institutional fingerprint when an organisation enforces
shared templates or review checklists.

\paragraph{Stage 2: Tool-assisted generation and templating.}
Most production-grade rules are not written from scratch.
Generators such as \texttt{yarGen}~\cite{yarGen} parse a malware
sample, filter candidate strings against a goodware corpus, and
emit a templated skeleton that the analyst then refines. The
resulting tool signature is largely independent of the targeted
malware: a \texttt{yarGen}-generated ransomware rule and a
\texttt{yarGen}-generated RAT rule share the same skeleton,
naming scheme, and condition structure, so any classifier
that learns the skeleton attributes rules to the generator
regardless of their target.

\paragraph{Stage 3: Normalisation and quality gates.}
Before acceptance into a repository, rules typically pass through
linters and reformatters that enforce uniform indentation,
consistent operator spacing, canonical hex casing, and prescribed
keyword ordering. These gates replace \emph{individual}
variability with \emph{repository-wide} uniformity: every rule
that passes through the same gate emerges structurally similar to
every other rule from that gate. 

\paragraph{Stage 4: Translation and backend adaptation.}
Rules are frequently rewritten for specific backends
(VirusTotal Livehunt, endpoint agents, sandbox pipelines):
regex expansion, PE-module substitution, condition-guard
insertion, module renaming. Because organisations adopt consistent
backends, these rewrites are repeated identically across many
rules, imprinting a backend-specific pattern on top of stages 1--3.

\paragraph{Stage 5: Distribution and sharing.}
The rule is published through a repository, ISAC feed, or
bilateral exchange, typically after \texttt{meta:}-block
sanitisation. The artefacts introduced by stages 1--4 are
preserved \emph{inside} the rule body, clause ordering, condition
nesting, naming conventions, boilerplate, because they are
required for the rule to remain syntactically valid. Metadata
removal thus leaves the stylometric surface intact.

\paragraph{Implications.}
The cumulative effect of these five stages is that every shared YARA
rule carries a layered fingerprint: individual authorial habits at
stage 1, tool signatures at stage 2, repository conventions at stage 3,
backend adaptations at stage 4, and residual formatting at stage 5. In practice, some of those dimension cannot be separated: human and tools working together at phase 1 and 2, in a co-authoring manner. On the other extremity of the pipeline, stage 3, 4 and 5 are hardly distinguishable: all those steps are sub-parts of the production and sharing rules policy of an organization, and thus, are not independent dimensions. We therefore hereafter only two dimensions, that we name hereafter \textit{author} and \textit{repository}. \textit{Authors} groups human authors and tool they used, and \textit{repository} groups the end of the pipeline. In addition to those dimensions, the malware family the YARA rules detect is also a dimension of the fingerprint\footnote{We determined empirically that malware alone does not carry enough information, while the family it belongs to is reattributable well above chance.}.   
The central question of this paper is whether these layers are strong
enough, individually or in combination, to re-identify the producer of
an anonymised rule. As any of those layers signature cand drift with time, we also studied if such a drift is identifiable, using methods and techiques coming from stylometric field.  Section~\ref{sec:related-work} reviews the stylometric and threat-intelligence literature related to our work.

\section{Related Work}
\label{sec:related-work}

\subsection{Source Code Authorship Attribution}

Source code authorship attribution (SCAA) rests on the principle of stylometry: that programmers exhibit a unique and distinguishable ``fingerprint'' in their coding style, analogous to a writer's literary style~\cite{burrows2013source}.
Authorship attribution holds significant importance in cybersecurity for attributing malware and in software forensics for addressing plagiarism, intellectual property theft, and copyright disputes~\cite{burrows2013source,Alalawi}.

\paragraph{Feature engineering era.}
Early work focused on engineering features from code text, divided into lexical and layout categories.
Frantzeskou et al.\ demonstrated that byte-level n-grams could build effective ``author profiles,'' proving the approach viable even with limited and comment-stripped code samples~\cite{frantzeskou2007source}.
A significant advancement came from Caliskan-Islam et al., who introduced features derived from Abstract Syntax Trees (ASTs).
By combining structural features with lexical and layout features, they used a Random Forest classifier to de-anonymise programmers at scale, firmly establishing the privacy and security implications of SCAA in public repositories~\cite{caliskan2015anonymizing}.

\paragraph{Deep learning transition.}
As the field progressed, research transitioned from conventional machine learning (SVMs, Random Forests) to deep learning.
Alsulami et al.\ were among the first to show that LSTM networks trained on AST-derived features could significantly surpass models with manually crafted features~\cite{alsulami2022sota}.
Comparative studies validated this pattern:
Frankel and Ghosh assessed both logistic regression and deep learning comprising NLP-derived features alongside ASTs~\cite{9671332}, while Alalawi et al.\ compared SVM and RF with LSTM, RNN, and CNN architectures, finding that deep learning consistently delivered superior performance~\cite{Alalawi}.
Ullah et al.\ further advanced feature extraction by merging Program Dependence Graphs (PDGs) with deep learning to capture data and control-flow dependencies~\cite{furhandgdl}.

\paragraph{Transformer era.}
Recently, the field has adopted pre-trained transformer models.
Dipongkor et al.\ performed a comprehensive empirical investigation on models such as CodeBERT, finding these models not only highly effective but also considerably more resilient to adversarial attacks compared to traditional ML and DL methods~\cite{dipongkor2025}.
Alvarez-Fidalgo et al.\ introduced CLAVE, a Transformer Encoder employing contrastive learning for authorship verification~\cite{ALVAREZFIDALGO2025104005}.
Choi et al.\ investigated zero-shot prompting with general-purpose LLMs (GPT-4) for attribution~\cite{choi2025isecondsleveraginglarge}, though this approach faces validity concerns regarding data contamination, LLMs trained on open-source code may recall code-author pairs rather than learn genuine stylistic features.
Our approach, using fine-tuned CodeBERT with controlled train/test splits, avoids this confound.

This body of work traces a clear progression: from lexical n-grams to structural ASTs, and from traditional ML to transformers.
However, prior research has predominantly focused on general-purpose programming languages (C++, Java, Python) from academic contests or large-scale open-source projects. ~\cite{horvath2026bridgingbehavioralbiometricssource}
To the best of our knowledge, the specialised domain of cybersecurity detection rules (YARA, Sigma, Snort, \ldots) has not been systematically investigated for attribution risk.

\subsection{Threat Intelligence Sharing and Its Risks}

The sharing of cyber threat intelligence (CTI), including detection signatures, is a well-established practice supported by formal standards such as STIX and TAXII~\cite{wagner2019cyber}.
At the institutional level, ISACs provide trusted communities where sector-specific organisations exchange threat data using protocols like the Traffic Light Protocol (TLP)~\cite{enisa2017isac}.
In Europe, the NIS2 Directive has further reinforced this model by mandating cooperation and information sharing across critical infrastructure operators~\cite{nis2directive}.
National agencies such as France's ANSSI (through CERT-FR) and the US CISA actively publish detection rules as part of their threat advisories~\cite{certfr2021sandworm}.

Despite this widespread sharing, the privacy implications of sharing detection rules have received comparatively little attention~\cite{esteban2026miningyaraecosystemadhoc}.
The Equation Group analysis by Kaspersky~\cite{kaspersky2016equation} demonstrated that persistent code-level stylistic characteristics can identify advanced threat actors, suggesting a symmetric risk for defenders sharing their own detection logic.
Yet no prior work has systematically examined whether detection rules \emph{as authored artefacts}, as opposed to the malware they detect, carry exploitable fingerprints~\cite{yao2026smudgedfingerprintssystematicevaluation}.

\section{Problem Statement}
\label{sec:problem}

This section formalises the threat model previewed in Section~\ref{sec:introduction} and defines the attribution tasks we evaluate.

\paragraph{Notations.}
Let $\mathcal{R} = \{r_1, r_2, \ldots, r_N\}$ be a corpus of anonymised
YARA rules, where each $r_i$ has been stripped of its \texttt{meta:}
block, block comments, and tags (see Section~\ref{subsec:preprocessing}).
Each rule carries three ground-truth labels drawn from the
production pipeline of Section~\ref{sec:rule-production}: its individual
author $a_i \in \mathcal{A}$, its source repository $p_i \in \mathcal{P}$,
and the malware it targets $m_i \in \mathcal{M}$. An adversary
observes only the anonymised rule body $r_i$ and seeks to recover one
or more of these labels. Note that the author label from YARA actually embed two dimensions identified earlier: human author and tools, as we don't have access to the author tool setup. Some of the tools cite themselves as author in the corresponding field. Human authors are free to discard the tool's name and replace it by their name, or to keep them as co-author of the tool. In this later case, that represents more than a third of our dataset, we considered the tool as a co-author of the rule. 

To study how attribution risk evolves over time, we additionally
construct a time-restricted subset $\mathcal{R}_{\text{timed}}
\subset \mathcal{R}$ containing only rules authored during 2022--2025
in \texttt{Neo23x0/signature-base} (the only repository in our
corpus that exposes reliable per-rule authoring timestamps through
its Git history; see \S\ref{subsec:dataset}). Temporal drift is then operationalised as the
\emph{difference} in repository-attribution accuracy between
$\mathcal{R}$ and $\mathcal{R}_{\text{timed}}$, rather than as a
per-rule label. 

\paragraph{Threat model and attacker capabilities.}
We consider a \emph{passive} adversary operating in the collaborative YARA sharing ecosystem (public repositories, ISAC feeds, CERT advisories). The adversary does not inject, modify, or delete rules, and does not compromise the sharing infrastructure. Its objective is to \emph{de-anonymise the producer} of shared rules and, by doing so, infer sensitive information about the originating organisation. The adversary has access to (i) a public sample of anonymised rules with ground-truth labels, which it uses to train classifiers, and (ii) one unlabelled anonymised rule whose producer it seeks to identify. It does not have access to the original unredacted rules, repository commit histories, internal IDS telemetry, or the tools and templates used to generate the rules.  The adversary operates only on the shared rule artefact itself, that is, its post-anonymisation rule body (strings, condition, inline comments). We assume no side channels (network timing, platform identifiers, infrastructure logs) and no auxiliary metadata.

\paragraph{Attribution tasks.}
We define four attribution tasks over the label set and subsets above, each corresponding to one of the fingerprint dimensions (identified as candidates in section \ref{sec:rule-production}, with the exception stated earlier that author and tool signatures cannot be distinguished in our dataset)
introduced by the production pipeline:
\begin{enumerate}
    \item \textbf{Author attribution:} learn $f_a : r \to \mathcal{A}$ to recover the individual analyst who wrote the rule, trained on $\mathcal{R}_{\text{timed}}$.
    \item \textbf{Repository attribution (time-restricted):} learn $f_p : r \to \mathcal{P}$ on $\mathcal{R}_{\text{timed}}$.
    \item \textbf{Repository attribution (full history):} train the same classifier $f_p$ on the full corpus $\mathcal{R}$. Comparing the time-restricted and full-history variants isolates the effect of temporal drift in repository conventions.
    \item \textbf{Malware family classification:} learn $f_m : r \to \mathcal{M}$ to recover the threat category the rule targets. 
\end{enumerate}

\emph{Security properties at stake.} Following NIST SP~800-150's principle of \emph{limited attribution}~\cite{NIST.SP.800-150}, rule sharing should preserve two properties: \emph{producer anonymity} (a shared rule should not be attributable to its producer beyond chance) and \emph{unlinkability} (multiple shared rules should not be reliably associable with the same producer). Our attribution tasks test both: single-rule attribution violates producer anonymity, and the per-repository and per-family results violate unlinkability. 


\paragraph{Research questions.}
Three research questions are addressed in this paper about those attribution tasks:

\begin{description}
    \item[RQ1 (Feasibility):] Can each of the attribution tasks above
    be solved substantially above random baseline on anonymised YARA
    rules, and what are the best methods to do that? 
    \item[RQ2 (Temporal drift):] Does the accuracy of $f_p$ on
    $\mathcal{R}_{\text{timed}}$ differ from its accuracy on
    $\mathcal{R}$, and if so, how?
    \item[RQ3 (Style vs.\ content):] When author attribution is
    performed \emph{within} a single malware family (and optionally
    within a single repository), does the signal persist? If yes, the
    classifier is learning stylistic fingerprints rather than content.
\end{description}

To answer those questions, we run experiment using the 3 main state-of-the-art classification methods (N-grams, AST, and transformers) and a confound-controlled experiment to disentangle the different dimension influencing attribution. The methodology we use is given in the next section.
\section{Methodology}
\label{sec:methodology}

\subsection{Dataset Construction}
\label{subsec:dataset}

We collected YARA rules from three major public GitHub repositories:
{Neo23x0/signature-base} (maintained primarily by Florian
Roth), {reversinglabs-yara-rules} (ReversingLabs), and
{YARAHQ/yara-forge}. After parsing, deduplication, and
metadata stripping, the unified corpus contains 23{,}305 rules
spanning 67 authors and 3 repositories. Each rule carries author,
repository, and malware-family annotations.
All author, full-history repository, and malware-family experiments
train on this unified corpus
({yara4\_with\_repo.csv}). Task-specific class-size filters are
applied independently to this single source
(\S\ref{subsec:preprocessing}).

\paragraph{Time-restricted subset.}
For the timed repository-attribution task, we additionally use
\texttt{yara2.csv}, a subset comprising the 14{,}366 rules authored
during 2022--2025 in \texttt{Neo23x0/signature-base}, the only
repository in our corpus that exposes reliable per-rule authoring
timestamps through its Git history. Comparing full-history and
time-restricted repository attribution isolates the effect of
temporal drift in repository conventions (\S\ref{sec:temporal-discussion}).

\paragraph{Malware-family labels.}
Family labels were assigned via a two-step pipeline: the
\texttt{rule\_name} and \texttt{meta:} description fields were
extracted from each raw rule and submitted to a large language model
(Claude, Anthropic) with a structured prompt requesting classification
into a fixed taxonomy of 22 families (e.g., APT, Backdoor,
Ransomware, Webshell). Rules that could not be confidently categorised
were assigned the label \texttt{Other} and excluded from the
malware-family task, yielding 16 retained families. Note that the
\texttt{rule\_name} and \texttt{meta:} description fields are not
included in the rule text presented to classifiers; they are used
solely for label assignment and are stripped during preprocessing
(\S\ref{subsec:preprocessing}).
\subsection{Preprocessing and Anonymisation}
\label{subsec:preprocessing}

Each rule undergoes a two-stage preprocessing step designed to eliminate
explicit identifiers while preserving the stylometric surface.

\paragraph{Stripping.}
We remove the \texttt{meta:} block in its entirety, as it contains the
strongest explicit identifiers (author name, description, date, hash
references). Block comments (\texttt{/* \ldots\ */}) are also removed
because they frequently contain copyright headers or analyst signatures.
Rule-level tags (\textit{e.g.}, \texttt{rule Name : \textbf{tag}}) are
stripped for the same reason.

\paragraph{Retention.}
We deliberately retain the {strings:} section, the
{condition:} block, and any \emph{inline} comments. These three
elements constitute the primary stylometric surface of a YARA rule: the
choice of variable names, hexadecimal encoding preferences, string literal
formatting, and the structure of boolean conditions are all degrees of
freedom that analysts exercise differently. As established in
Section~\ref{sec:rule-production}, these are precisely the layers at
which both individual authorial habits and tool-induced fingerprints
are introduced and persist through the sharing pipeline.

\paragraph{Class-size filtering.}
For author attribution, we retain only authors with between 50 and
500 rules. The lower bound of 50 reflects the minimum number of
samples needed to obtain a meaningful 80/20 stratified split
(40 training rules / 10 validation rules per author); below this
threshold, validation estimates become dominated by sampling noise
and the model has insufficient signal to learn a stable stylistic
profile. The upper bound of 500 addresses a pronounced long-tail
imbalance in our corpus: a small number of prolific maintainers
(notably Florian Roth, who authored 56\% of the unfiltered dataset)
would otherwise dominate the class distribution, allowing the
classifier to achieve high apparent accuracy by learning a single
dominant style rather than discriminating among authors. Capping
at 500 rules per author yields a more balanced training set in
which macro-averaged F1 becomes a meaningful indicator of
per-author attribution risk rather than a reflection of the
majority class. For repository and malware-family tasks, we retain
all classes with at least 50 rules under the same minimum-signal
rationale; no upper cap is applied, as these tasks have
substantially fewer classes (3 and 16 respectively) and the
class-imbalance effect is less severe. An 80/20 stratified
train--validation split is applied uniformly across all tasks.

\subsection{Corpus-level artefacts and leakage controls}
\label{subsec:corpus-limitations}

YARA repositories are community resources in which rules are
frequently re-used, refined, or re-attributed across analysts. Our
extraction pipeline emits one CSV row per (rule, author) pair, where
authorship is taken from each rule's \texttt{meta} block. Two
properties of the upstream corpus and our pipeline interact in ways
that warrant explicit discussion.

\paragraph{Multi-attribution rules.} Some rules in the upstream
repositories appear with author strings that span multiple analysts. Our
pipeline treats each distinct comma-separated author token as a
separate label, which means a rule whose \texttt{meta} declares
two authors generates two CSV rows with identical rule bodies and
distinct labels. In our master corpus, 7{,}442 of 19{,}224 rows
share a \texttt{rule\_name} with at least one other row; 2{,}182
distinct rule names appear with byte-identical bodies under multiple
author labels. We retain these rows because they reflect the true
distribution of analyst attributions in published YARA corpora,
which is the population a downstream attacker would observe; removing
them would also remove a real signal about co-authorship patterns
in the threat-intelligence community.

\paragraph{Effect on stratified splits.} Where multi-attribution
rows exist, our 80/20 stratified split
(Section~\ref{subsec:preprocessing}) may place identical bodies under
different labels into both training and validation sets. This affects
the author-attribution task in Table~\ref{tab:main-results} most
directly; repository, malware-family, and timed repository tasks are
largely unaffected, as their labels are not author-derived. The
per-family confound-controlled experiments of
Section~\ref{sec:results-perfamily} train within fixed
\texttt{(family, repository)} subsets and substantially constrain
the duplication pattern.

\paragraph{Identifier-leakage ablation.} A second leakage channel,
distinct from row-level duplication, is whether attribution accuracy
is driven by analyst signatures, tool names, or URLs embedded in
inline \texttt{//} comments. We address this directly in
Section~\ref{subsec:ablation-comments} by retraining both the author
and timed-repository tasks on a corpus variant in which all inline
comments have been stripped. Accuracy changes by at most 2~pp and
moves in the direction opposite to the identifier-leak hypothesis
(stripping comments slightly improves author attribution rather than
degrading it), supporting the claim that the stylometric signal
resides in the rule body and condition logic rather than in
human-readable identifiers.

\subsection{Classification Methods}
\label{sec:methods}

We apply three complementary classifiers.

\paragraph{Lexical n-grams.}
Following Burrows et al.~\cite{burrows2013source}, we extract
character n-grams ($n=1,\ldots,6$) from raw rule text and
classify with a nearest-centroid classifier under Burrows'
Delta distance. This captures surface-level lexical habits.

\paragraph{AST features.}
Following Caliskan-Islam et al.~\cite{caliskan2015anonymizing},
we parse each rule into an abstract syntax tree and extract
node-type frequencies, tree depth, branching factors, and
node-type bigrams; classification uses a Random Forest with 300
estimators. This captures syntactic habits invariant to surface
formatting.

\paragraph{Fine-tuned CodeBERT.}
We fine-tune \texttt{microsoft/codebert-base}~\cite{feng2020codebert}
with a linear classification head, training one independent
single-task model per attribution task. Author, full-history
repository, and malware-family tasks train on the unified corpus;
the timed repository task trains on the Neo23x0 2022--2025 subset.
Rule text is tokenised and truncated to 512 tokens; training uses
AdamW (lr~$2\times10^{-5}$, batch~16) with early stopping
(patience~3) on validation accuracy. Experiments ran on
PyTorch~1.10.2 (CUDA~11.1) and HuggingFace Transformers~4.30.0;
the full environment is in the released artefact bundle.

All experiments were run with PyTorch~1.10.2 (CUDA~11.1) and HuggingFace Transformers~4.30.0; the complete environment specification is included in the released artifact bundle.
\subsection{Stylometric Fingerprint Dimensions}
\label{sec:axes}

We evaluate each method to perform the attribution tasks defined in section \ref{sec:problem}, with our dataset, i.e.: 

\begin{enumerate}
    \item \textbf{Author Attribution (17 classes):} Predict the individual author of a rule from the unified corpus, after applying the 50--500 rules-per-author class-size filter (\S\ref{subsec:preprocessing}).
    \item \textbf{Repository Attribution -- Timed} (3 classes): Predict the source repository from the same time-restricted subset.
    \item \textbf{Repository Attribution -- Full} (3 classes): The same repository prediction task on the full time-range dataset.
    \item \textbf{Malware Family Classification} (16 classes): Predict the malware family a rule targets.
\end{enumerate}

Comparing Repository Attribution (Full) against Repository Attribution (Timed) isolates the effect of temporal drift in repository conventions: if conventions have become more standardised over the 2022--2025 window, we expect the timed subset to yield higher accuracy.
\subsection{Evaluation Metrics}
\label{subsec:metrics}

Given the severe class imbalance across our datasets (see
Section~\ref{subsec:dataset}), reporting a single accuracy figure would
obscure per-class performance. We therefore report three metrics
throughout the paper.

\paragraph{Accuracy.}
The fraction of validation rules whose predicted label matches the
ground truth. Accuracy is the simplest summary but is dominated by
the majority class under imbalance.

\paragraph{Macro-averaged F1 (M-F1).}
F1 is computed independently for each class and then averaged with
equal weight, regardless of class size:
\begin{equation}
    \mathrm{M\text{-}F1} = \frac{1}{|C|} \sum_{c \in C} \mathrm{F1}_c.
\end{equation}
Because every class contributes equally, M-F1 is sensitive to
performance on minority classes. From an OPSEC standpoint this is
the more diagnostic metric: a single attributable author represents
a concrete re-identification risk regardless of how few rules they
have authored.

\paragraph{Weighted F1 (W-F1).}
Per-class F1 scores are averaged weighted by the number of samples
per class:
\begin{equation}
    \mathrm{W\text{-}F1} = \sum_{c \in C}
        \frac{|S_c|}{|S_{\text{total}}|} \, \mathrm{F1}_c.
\end{equation}
W-F1 reflects the expected outcome for a randomly sampled rule from
the corpus and is dominated by the most prolific classes.

\paragraph{Random baseline.}
For a classification task with $K$ equally likely classes, the
random-guessing baseline accuracy is $1/K$. For author attribution
within a malware family containing $K$ authors, the baseline is
$1/K$; for 3-way repository attribution it is $33.3\%$; for 16-way
malware classification it is $6.25\%$. We use this baseline as the
reference point for the attribution-verdict thresholds defined in
Section~\ref{subsec:verdicts}.

\paragraph{Reporting convention.}
In all tables we report accuracy, M-F1, and W-F1 on the held-out
20\% validation split, stratified by the target label. The gap
between M-F1 and W-F1 provides diagnostic insight into whether a
method relies on majority-class performance or generalises uniformly
across classes.
\subsection{Per-Family Confound Control}
\label{sec:per-family}

A key concern in author attribution of detection rules is content confounding: authors who specialise in specific malware families may be identifiable by the \emph{content} they detect rather than their \emph{writing style}.
To disentangle these effects, we design a per-family experiment with two tiers:

\paragraph{Tier 1: Per-Family.}
For each malware family with 3 or more authors, each having at least 20 rules, we train an independent CodeBERT classifier. It predicts the author within that family only.
If attribution succeeds, the signal must come from style rather than content, since all rules in the training set target the same malware.

\paragraph{Tier 2: Per-Family $\times$ Per-Repository.}
We further partition by repository, training author classifiers within each (family, repository) pair.
This eliminates organisational coding standards as an alternative explanation, isolating individual stylistic fingerprints.
\paragraph{Attribution verdicts.}
\label{subsec:verdicts}
We classify each per-family experiment by the ratio of its observed
accuracy to the random baseline $1/K$, where $K$ is the number of
candidate authors in that experiment. Three verdict bands are
defined:
\begin{description}
    \item[Attributable] ($\mathrm{acc} > 2 \times 1/K$): the classifier
    substantially exceeds random guessing, indicating that stylistic
    fingerprints are strong enough to pose a concrete
    re-identification risk even under content or organisational
    control.
    \item[Partially resilient] ($1.5 \times 1/K < \mathrm{acc}
    \leq 2 \times 1/K$): the classifier outperforms random guessing
    but not decisively; fingerprints exist but are attenuated by
    template use or style overlap.
    \item[Resilient] ($\mathrm{acc} \leq 1.5 \times 1/K$): the
    classifier is at or near random performance, indicating that
    the combination of controlled content, organisational
    standardisation, and limited stylistic variation suppresses
    identifiable fingerprints.
\end{description}
The choice of $2\times$ as the ``clearly attributable'' threshold
reflects the OPSEC interpretation: if an adversary's posterior
belief about the author of a rule is at least twice the uniform
prior, the rule has leaked meaningful information about its
producer. The $1.5\times$ intermediate band captures cases where
the leak is statistically detectable but operationally marginal.
\section{Results}
\label{sec:results}

\subsection{Multi-Method Comparison}
\label{sec:main-results}

Table~\ref{tab:main-results} presents the complete results for the three methods we used applied to the four fingerprint dimensions identified.
All metrics are reported on held-out validation sets (20\% stratified split).

\begin{table*}[t]
\caption{Multi-method, multi-axis attribution results. Accuracy (\%), macro-averaged F1 (M-F1), and weighted F1 (W-F1) are reported. Best result per task in \textit{bold}.}
\label{tab:main-results}
\centering
\begin{tabular}{llccccccccc}
\toprule
& & \multicolumn{3}{c}{\textbf{N-grams}} & \multicolumn{3}{c}{\textbf{AST}} & \multicolumn{3}{c}{\textbf{CodeBERT}} \\
\cmidrule(lr){3-5} \cmidrule(lr){6-8} \cmidrule(lr){9-11}
\textbf{Task} & \textbf{Classes} & Acc & M-F1 & W-F1 & Acc & M-F1 & W-F1 & Acc & M-F1 & W-F1 \\
\midrule
Author & 17 & 58 & 0.60 & 0.57 & 52 & 0.54 & 0.51 & \textbf{76} & \textbf{0.76} & \textbf{0.75} \\
Repo (Timed) & 3 & \textbf{99} & \textbf{0.99} & \textbf{0.99} & 98 & 0.98 & 0.98 & \textbf{99} & \textbf{0.99} & \textbf{0.99} \\
Repo (Full) & 3 & 81 & 0.85 & 0.79 & 83 & 0.87 & 0.83 & \textbf{90} & \textbf{0.93} & \textbf{0.91} \\
Malware & 16 & 93 & 0.79 & 0.92 & 89 & 0.65 & 0.88 & \textbf{95} & \textbf{0.93} & \textbf{0.95} \\
\bottomrule
\end{tabular}
\end{table*}

\begin{figure}[t]
\centering
\includegraphics[width=\columnwidth]{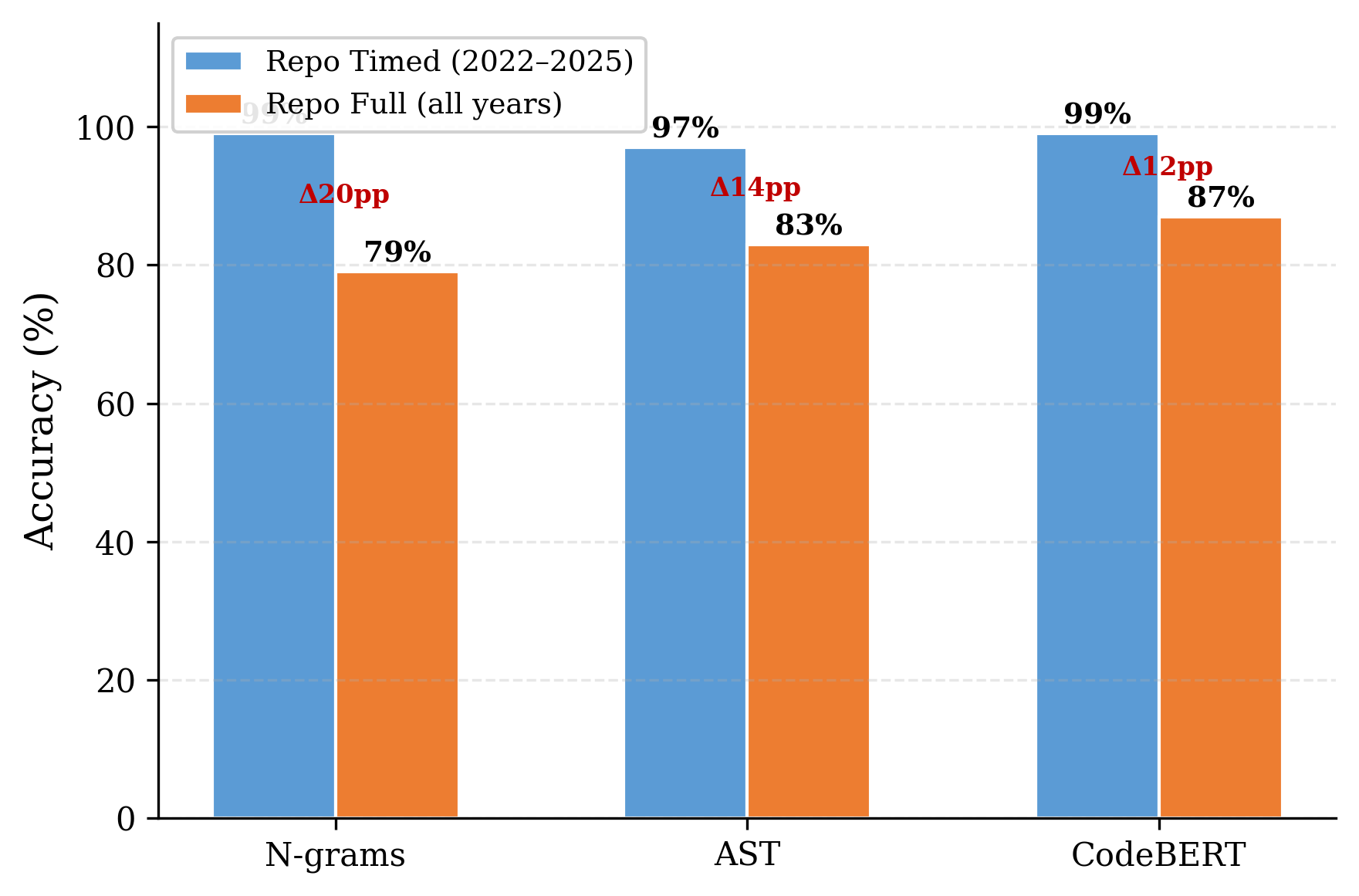}
\caption{Temporal convergence effect: repository attribution accuracy on the 2022--2025 subset vs.\ full history.}
\Description{Grouped bar chart showing Repo Timed versus Repo Full accuracy for three methods. Gaps are 20pp for N-grams, 14pp for AST, and 12pp for CodeBERT.}
\label{fig:temporal-convergence}
\end{figure}

Second, \textbf{repository attribution on the timed subset approaches ceiling performance} (99\% for both n-grams and CodeBERT), while the same task on the full dataset drops to 81--90\%.
This 9--18 percentage point gap is consistent with repository-specific conventions converging over the 2022--2025 period, which would make rules from this window more distinctively fingerprinted.
Figure~\ref{fig:temporal-convergence} isolates this temporal effect. We examine alternative explanations for the gap in Section~\ref{sec:temporal-discussion}, such as virus drift.

Two caveats constrain the interpretation of this result. First, only
Neo23x0/signature-base exposes reliable per-rule authoring timestamps
through its Git history; ReversingLabs and YaraHQ/yara-forge provide
rules without dependable temporal metadata. The ``timed'' subset is
therefore effectively a Neo23x0-weighted slice of the corpus, and the
gap should be read as a \emph{within-repository} observation rather
than a cross-ecosystem trend. Second, the 2022--2025 window spans only
four calendar years. We therefore treat this result as preliminary
evidence that temporal drift in repository fingerprints is
\emph{measurable}, not as a definitive characterisation of how
repository conventions evolve.

Third, \textbf{n-grams outperform AST features on content-heavy tasks} (malware: 93\% vs.\ 89\%).
This is expected: malware-specific string patterns (e.g., hexadecimal signatures) are captured by lexical n-grams but not by structural AST features.
Conversely, AST features slightly outperform n-grams on the full repository task (83\% vs.\ 81\%), suggesting that structural conventions (e.g., condition complexity) are more stable across time than surface-level formatting.

\subsection{Per-Family Author Attribution (Confound Control)}
\label{sec:results-perfamily}
To estimate if author attribution good results may be due to other fingerprints dimensions, we performed authors attribution by splitting datasets by malware families, using only rules targeting a specific family for both training and testing (tier 1).   
Table~\ref{tab:tier1} shows these author attribution results within each malware family.

\begin{table}[t]
\caption{Per-family author attribution with CodeBERT. Random baseline is $\frac{1}{\text{authors}}$.}
\label{tab:tier1}
\centering
\begin{tabular}{lrrrl}
\toprule
\textbf{Family} & \textbf{Authors} & \textbf{Rules} & \textbf{Acc (\%)} & \textbf{Verdict} \\
\midrule
Ransomware & 3 & 358 & 91.7 & Attributable \\
Backdoor & 5 & 310 & 88.7 & Attributable \\
Exploit & 4 & 339 & 88.2 & Attributable \\
APT & 9 & 514 & 77.7 & Attributable \\
RAT & 3 & 285 & 70.2 & Attributable \\
Hacktool & 19 & 2{,}144 & 61.3 & Partial \\
Webshell & 4 & 632 & 44.1 & Resilient \\
\midrule
\multicolumn{3}{l}{\textit{Mean (7 families)}} & 74.6 & --- \\
\bottomrule
\end{tabular}
\end{table}

\begin{figure}[t]
\centering
\includegraphics[width=\columnwidth]{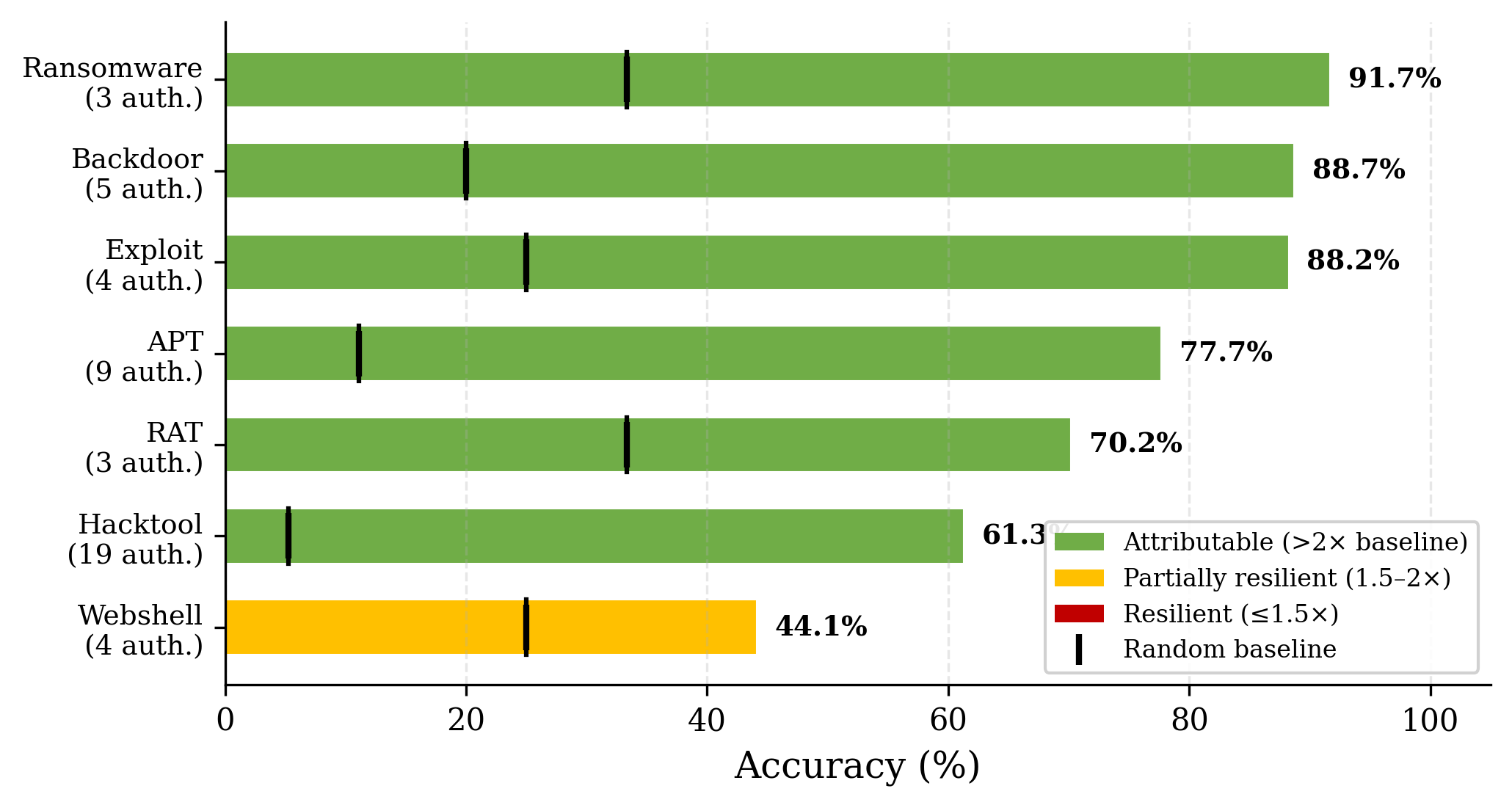}
\caption{Tier~1: Per-family author attribution accuracy.}
\Description{Horizontal bar chart showing 7 malware families ordered by accuracy. Ransomware (91.7\%), Backdoor (88.7\%), and Exploit (88.2\%) are attributable. Webshell (44.1\%) is resilient.}
\label{fig:tier1-bar}
\end{figure}

Five of seven tested families are classified as \emph{attributable}, with Ransomware achieving 91.7\% accuracy across 3 authors (random baseline: 33\%).
Only Webshell exhibits resilience to attribution (44.1\% with 4 authors, baseline 25\%), likely because webshell rules follow highly standardized structural templates that suppress individual variation.
Hacktool, the largest and most diverse family (19 authors, 2{,}144 rules), falls in the \emph{partially resilient} category at 61.3\% (baseline 5.3\%), still far above chance.
Figure~\ref{fig:tier1-bar} visualizes these results against their random baselines.

\begin{table}[t]
\caption{Tier~2: Per-family $\times$ per-repository author attribution. Controls for both content and organizational conventions.}
\label{tab:tier2}
\centering
\begin{tabular}{llrrl}
\toprule
\textbf{Family $\times$ Repo} & \textbf{Auth.} & \textbf{Rules} & \textbf{Acc (\%)} & \textbf{Verdict} \\
\midrule
Exploit $\times$ Neo23x0 & 3 & 254 & 94.1 & Attrib. \\
APT $\times$ YaraHQ & 3 & 303 & 80.3 & Attrib. \\
APT $\times$ Neo23x0 & 5 & 324 & 78.5 & Attrib. \\
Webshell $\times$ Neo23x0 & 3 & 431 & 77.0 & Attrib. \\
Backdoor $\times$ YaraHQ & 3 & 121 & 68.0 & Partial \\
Hacktool $\times$ Neo23x0 & 8 & 937 & 67.0 & Partial \\
Hacktool $\times$ YaraHQ & 13 & 1{,}497 & 54.7 & Partial \\
Webshell $\times$ YaraHQ & 3 & 600 & 33.3 & Resilient \\
\bottomrule
\end{tabular}
\end{table}

\begin{figure}[t]
\centering
\includegraphics[width=\columnwidth]{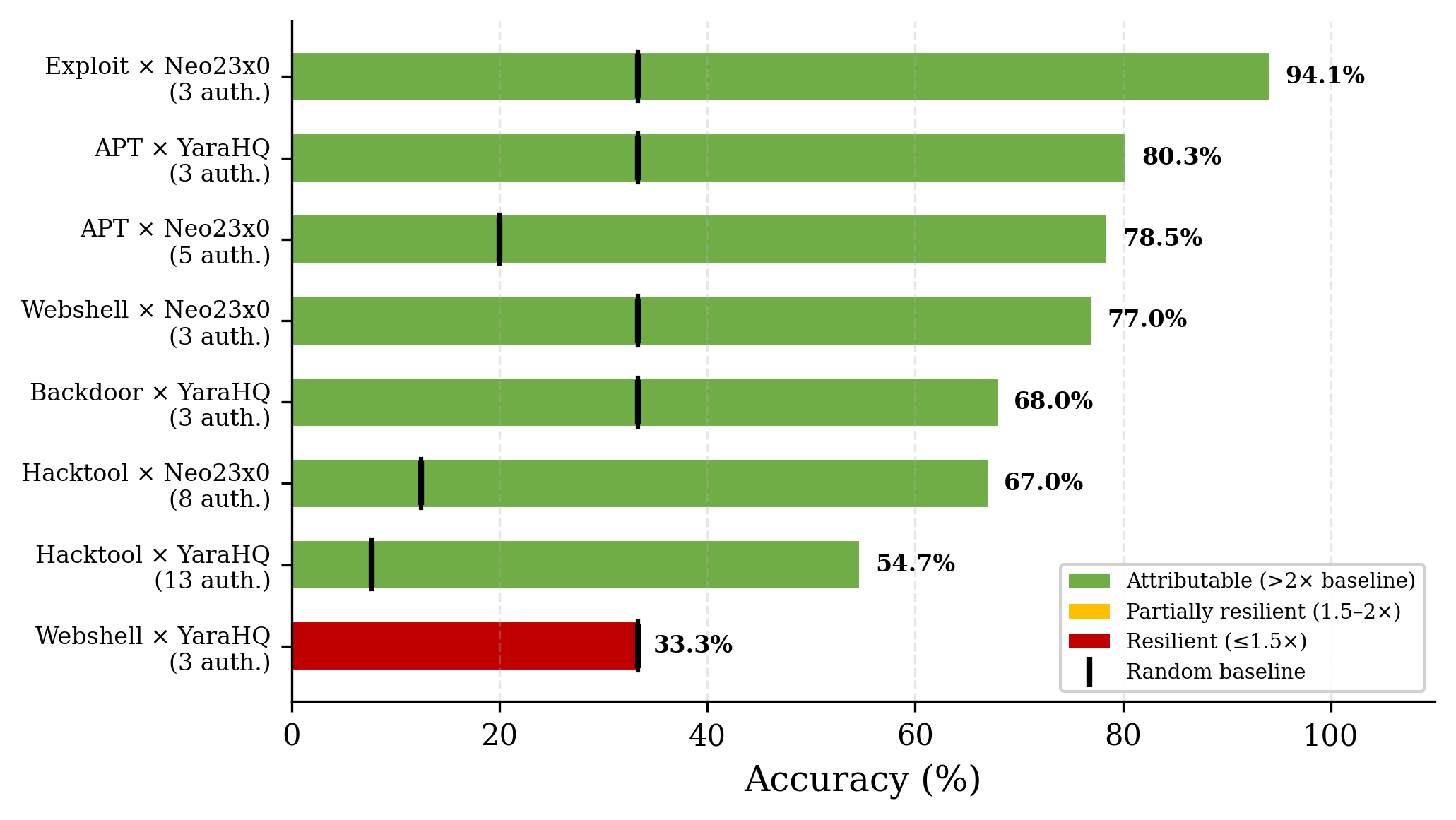}
\caption{Tier~2: Per-family $\times$ per-repository author attribution.}
\Description{Horizontal bar chart showing 8 family-repo combinations. Exploit×Neo23x0 (94.1\%) is the most attributable. Webshell×YaraHQ (33.3\%) collapses to random chance.}
\label{fig:tier2-bar}
\end{figure}

Table~\ref{tab:tier2} shows Tier~2 results, i.e. results for each family and repository couple.
Even after controlling for both malware family \emph{and} repository, four of eight experiments remain attributable.
Notably, Webshell~$\times$~Neo23x0 is attributable at 77.0\% while Webshell~$\times$~YaraHQ collapses to chance (33.3\%), suggesting that YaraHQ's webshell rules follow such rigid templates that individual authorship signals are completely suppressed.
This contrast provides direct evidence that organizational style conventions can either amplify or suppress individual fingerprints.
Figure~\ref{fig:tier2-bar} presents these results visually.

\label{sec:varying-n}

The fixed per-author cap used in Section~\ref{sec:results-perfamily}
leaves one question unanswered: how does the amount of training data
available per author shape what the classifier actually learns? To
probe this, we repeat the Hacktool per-family experiment while sweeping
a fixed per-author rule cap $N \in \{20, 80, 100, 200, 400\}$ across
both tiers. Varying $N$ has two coupled effects. First, it changes the
number of training samples available per author. Second, it implicitly
changes the set of authors who qualify for inclusion, since authors
with fewer than $N$ rules are excluded at each cap. This design
exposes how fingerprint signal concentrates as the per-author corpus
grows and as the author set narrows to the most prolific contributors.

\begin{table}[t]
\caption{Hacktool per-family accuracy as the per-author rule cap $N$
varies. $K$ is the number of qualifying authors at each cap; random
baseline is $1/K$.}
\label{tab:varying-n}
\centering
\small
\begin{tabular}{lrrrrr}
\toprule
\textbf{Setting} & $N{=}20$ & $N{=}80$ & $N{=}100$ & $N{=}200$ & $N{=}400$ \\
\midrule
T1: Hacktool                & 50.0\%  & 57.5\%  & 60.6\%  & 63.2\%  & 65.0\%  \\
\quad ($K$ authors)         & (19)    & (10)    & (8)     & (7)     & (6)     \\
T2: Hacktool$\times$Neo23x0 & 68.8\%  & 53.1\%  & 53.8\%  & ---      & ---     \\
\quad ($K$ authors)         & (8)     & (4)     & (4)     &         &         \\
T2: Hacktool$\times$YaraHQ  & 48.1\%  & 53.9\%  & 62.5\%  & 60.0\%  & 49.7\%  \\
\quad ($K$ authors)         & (13)    & (8)     & (6)     & (5)     & (4)     \\
\bottomrule
\end{tabular}
\end{table}

\paragraph{Accuracy trends.}
In the unrestricted Tier 1 setting, raw accuracy rises monotonically
from 50.0\% at $N{=}20$ (19 authors, baseline 5.3\%) to 65.0\% at
$N{=}400$ (6 authors, baseline 16.7\%). Because the baseline also
rises as authors are filtered out, raw accuracy alone is misleading;
the accuracy-to-baseline ratio is the meaningful quantity. That ratio
remains consistently in the \emph{attributable} band: $9.4\times$ at
$N{=}20$, falling to $3.9\times$ at $N{=}400$. The downward trend in
this ratio reflects the narrowing of the author set to the most
prolific, and therefore most homogeneously ``professional'', contributors,
whose styles are harder to separate. The Tier 2 settings show less
regular behaviour because the coupled filtering is more aggressive:
for Hacktool$\times$Neo23x0, only 4 authors retain at least 80 rules,
which mechanically caps achievable accuracy at the level of the
closest pair. This confirms that the fingerprinting signal observed
in our headline results is robust to sample-size choices and is not an
artefact of any single cap.

\begin{figure*}[t]
\centering
\includegraphics[width=\linewidth]{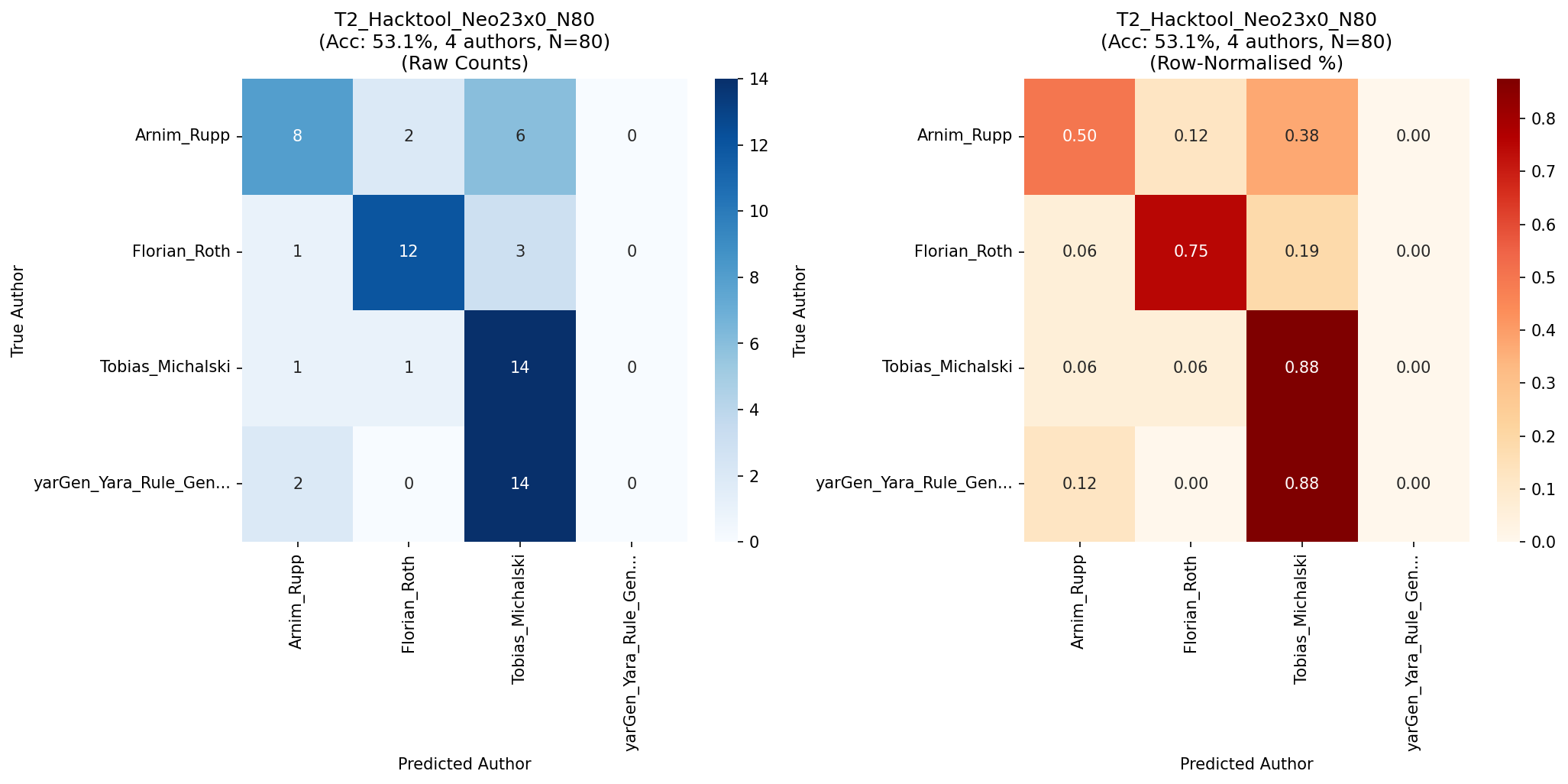}
\caption{Tier 2 Hacktool$\times$Neo23x0 confusion matrix at $N{=}80$
(row-normalised). }
\Description{88\% of \texttt{yarGen}-generated rules are
classified as Tobias Michalski. This directional confusion reflects a
real institutional link: Michalski is a Nextron Systems analyst, and
Nextron maintains the \texttt{yarGen} tool.}
\label{fig:cm-t2-hacktool-neo23x0-n80}
\end{figure*}

\paragraph{An unexpected institutional signal.}
More striking than the accuracy trend is a qualitative pattern that
emerges from the confusion matrices themselves. Across every Tier 2
Hacktool$\times$Neo23x0 run, the classifier systematically confuses
\texttt{yarGen}-generated rules with those written by Tobias Michalski (an employee of Nextron Systems, Nextron Systems being the company developping YarGen):
88\% of yarGen-labelled rules are predicted as Michalski at $N{=}80$
(Figure~\ref{fig:cm-t2-hacktool-neo23x0-n80}), 85\% at $N{=}100$, and
a similar directional pattern (44\% yarGen$\to$Michalski) persists in
Tier 1 at $N{=}80$. The confusion is directional, yarGen is predicted
as Michalski far more often than the reverse, and is stable across
caps. It is not the kind of symmetric error one would expect from
two genuinely similar authors; it has the profile of a many-to-one
mapping in which the ``yarGen'' class is being absorbed into the
``Michalski'' class. 

\paragraph{Implication: institutional is easier than individual.}
This finding extends the attribution story in two ways. First, it
demonstrates that the stylometric signal does not always decompose
cleanly into ``human author'' versus ``tool fingerprint.'' When
an analyst and a tool share a production environment, their
fingerprints \emph{converge}, and any classifier trained to separate
them will instead recover the signature of the couple formed by them. 

\subsection{Institutional Co-authorship in the Nextron Pipeline}
\label{subsec:institutional}

The varying-$N$ confusion described above admits a more concrete
explanation than stylometric similarity: the upstream
{Neo23x0/signature-base} repository \emph{declares} the
institutional coupling in its \texttt{meta} blocks. Of the 174
rules attributed to Tobias Michalski in
\texttt{Hacktool$\times$Neo23x0}, all 174 share an identical
\texttt{rule\_name} with rules attributed to \emph{yarGen}, and
174 with rules attributed to Florian Roth. Inspection shows the
bodies are byte-identical (Table~\ref{tab:nextron-overlap}). The
same content is published under multiple author labels, most
commonly because rules originally generated by \emph{yarGen} are
subsequently curated and re-attributed by analysts at the same
organisation. Our extraction pipeline
(Section~\ref{subsec:corpus-limitations}) preserves each declared
attribution as a separate row.

\begin{table}[t]
\centering
\caption{Co-attribution overlap among Nextron-affiliated authors
in \texttt{Hacktool\,×\,Neo23x0}.}
\Description{Each row reports the number of distinct
\texttt{rule\_name} values whose body appears under both authors
with $\geq 90\%$ Jaccard similarity over character 5-gram
shingles. All 174 Michalski-yarGen pairs are byte-identical;
edge counts in the other two pairs (172) reflect a small number
of upstream variants.}
\label{tab:nextron-overlap}
\begin{tabular}{lc}
\toprule
Author pair & Co-attributed rules \\
\midrule
Michalski $\leftrightarrow$ \emph{yarGen}     & 174 \\
Florian Roth $\leftrightarrow$ Michalski      & 172 \\
Florian Roth $\leftrightarrow$ \emph{yarGen}  & 172 \\
\bottomrule
\end{tabular}
\end{table}

The classifier's 88\% Michalski$\to$\emph{yarGen} confusion is
therefore not the discovery of a hidden institutional voice; it
is the recovery of a multi-attribution pattern visible to anyone
reading the upstream meta blocks. The two views, model confusion
and direct co-attribution counts, agree, which strengthens rather
than weakens the institutional-attribution finding: the coupling
between Nextron's tooling and its named analysts is robust enough
to be recovered by either method independently.


\paragraph{Generalisability.} The Nextron triad is the most
prominent example in our corpus but not unique: across the
unified corpus, 3{,}514 distinct rule names appear under multiple
author labels, of which 2{,}182 are byte-identical across
attributions (Section~\ref{subsec:corpus-limitations}). The
co-authorship signal we describe is a structural property of how
analyst-authored and tool-generated rules co-exist in publicly
shared corpora.

\section{Discussion}
\label{sec:discussion}

\subsection{What Drives Attribution}

Our results establish a clear hierarchy of attribution signal strength across fingerprint dimensions in our dataset:
\emph{repository} $>$ \emph{malware content} $>$ \emph{individual author} $>$ \emph{temporal period}.
Repository origin is the strongest signal (up to 99\%), consistent with organisational coding standards, naming conventions, boilerplate structure, condition templates, acting as a collective fingerprint.
Individual author attribution, while lower in absolute terms (76\%), is remarkable given that it succeeds even within a single malware family (Tier~1 mean: 74.6\%) and within a single family-repository pair (Tier~2: four attributable pairs).
The consistency across methods reinforces this hierarchy: even the weakest method (AST) achieves 83\% on repository attribution and 89\% on malware classification, confirming that these signals are robust and not an artefact of any single feature representation.

A key insight from our per-author analysis is that attribution vulnerability is not a simple function of data volume. The varying-$N$ sweep in Section~\ref{sec:varying-n} shows that narrowing the author set to only the most prolific contributors \emph{reduces}
the accuracy-to-baseline ratio from $9.4\times$ at $N{=}20$ to $3.9\times$ at $N{=}400$: when only high-volume contributors remain, their styles are more homogeneous and therefore harder to distinguish. This is the opposite of the intuition that ``more data
always helps the attacker'' and has direct implications for contributor OPSEC, a prolific author who conforms to community conventions may paradoxically be harder to re-identify than a low-volume author with an idiosyncratic style.

The varying-$N$ analysis in Section~\ref{sec:varying-n} further refines this hierarchy: when tool-induced and individual fingerprints originate from the same organisational  pipeline, they are not separable by a classifier, and the resulting signal is best
understood as an \emph{institutional} fingerprint rather than as either an individual or a tool one.

\subsection{Temporal Drift: A Premise, Not a Conclusion}
\label{sec:temporal-discussion}
The nine- to twenty-percentage-point accuracy gap between the timed and full-history repository tasks is consistent across all three methods, and the direction of the gap is the same in every case: the more recent the rules, the more attributable they become. Taken at face value, this is the signature one would expect if repository conventions were standardising over time, with newer rules passing through more uniform linting, templating, and review pipelines than older ones.

However, our data cannot establish this conclusion on its own. The temporal signal rests on a single repository, Neo23x0/signature-base, the only source in our corpus with reliable per-rule timestamps, and on a four-year window that is short relative to the decade-long
history of public YARA sharing. The observed effect could equally
reflect a Neo23x0-specific editorial shift, a change in its
contributor base, or a change in what kinds of threats the community
has prioritised during 2022--2025. We cannot, from our current
experiments, distinguish between these hypotheses.

We therefore present the temporal result as a \emph{premise} for
future work rather than as a finding. Two extensions are needed to
sharpen it. First, the analysis should be replicated on repositories
with richer temporal metadata, ideally spanning at least a decade,
and with contributor sets that are stable across the window. Second,
the attribution task itself should be decoupled from the filtering
effect: controlling for the set of authors who are active in both
sub-periods would isolate true convergence from a change in
contributor composition.

The method-specific variation in the gap (18\,pp for n-grams, 15\,pp
for AST, 9\,pp for CodeBERT) is nonetheless suggestive. If the effect
were purely compositional it would be roughly uniform across methods;
the fact that it is largest for surface-level features and smallest
for deep contextual embeddings is consistent with the hypothesis that
cosmetic standardisation has advanced faster than deeper structural
convergence. This pattern is worth reporting precisely because it
motivates the longer-horizon study needed to test it. For the
purposes of the present paper, the temporal result should be
understood as an invitation to that study, not as a completed one.

\subsection{The Content Confound and Its Resolution}

The comparison between aggregate-level malware classification (95\% for CodeBERT) and per-family author attribution (mean 74.6\%) reveals the extent of content confounding in naive attribution setups.
The 20-percentage-point drop when content is controlled confirms that a substantial portion, but not all, of the attribution signal in the unconstrained setting is driven by content rather than style.
A classifier trained on the full dataset without family controls would conflate ``Author X writes ransomware'' with genuine stylistic attribution.
Our per-family design eliminates this confound: within a single family, all rules target the same malware, so any above-chance accuracy must derive from stylistic differences.

The Webshell~$\times$~YaraHQ result (33.3\% = random chance) illustrates the limiting case: when organisational templates fully standardise rule structure, individual fingerprints vanish.
Conversely, Ransomware (91.7\%) and Exploit~$\times$~Neo23x0 (94.1\%) show that where templates are less rigid, authorship signals are strong and persistent.
\subsection{Ablation: Inline Comments}
\label{subsec:ablation-comments}

Because §\ref{subsec:preprocessing} deliberately retains inline \texttt{//} comments as part of the stylometric surface, a natural concern is whether attribution is driven by rule-body style or by analyst signatures, tool names, or URLs that may appear inside comments. We therefore rerun the two primary tasks on a variant of the corpus in which every inline comment has been removed, keeping the rest of the preprocessing pipeline identical.

Table~\ref{tab:ablation-inline} reports the comparison. Accuracy changes by at most 2 percentage points in either direction, and does so in the direction opposite to what an identifier-leak hypothesis would predict: stripping comments marginally \emph{improves} author attribution rather than degrading it. We conclude that the stylometric signal resides in the retained strings and condition blocks, not in analyst signatures or identifiers embedded in inline comments. Removing inline comments altogether would therefore be a defensible preprocessing choice; we retain them in the released corpus to preserve maximum fidelity to the shared-rule artefact as observed in practice.

\begin{table}[t]
\centering
\caption{Effect of removing inline \texttt{//} comments on
attribution accuracy. Accuracy is the best-epoch validation value.
Baseline values correspond to Table~\ref{tab:main-results}.}
\label{tab:ablation-inline}
\begin{tabular}{lccc}
\toprule
Task                          & Baseline & --inline & $\Delta$ \\
\midrule
Author (17 classes)           & 76.4    & 78.4    & $+2.0$ \\
Repository, timed (3 classes) & 99.2    & 99.3    & $+0.1$ \\
\bottomrule
\end{tabular}
\end{table}
\subsection{Macro vs.\ Weighted F1: Reading the Results}

Given the severe class imbalance across our datasets, we report both macro-averaged and weighted F1 scores.
Macro F1 gives equal weight to every class regardless of size, making it critical for judging performance on minority classes, a single attributable author constitutes a real OPSEC risk regardless of how many rules they have written.
Weighted F1 reflects the expected outcome for a randomly sampled rule.

The gap between the two metrics varies systematically across methods and provides diagnostic insight into \emph{what type} of signal each method relies on.
For the malware task, CodeBERT achieves 0.95 weighted F1 and 0.93 macro F1, a narrow gap indicating strong performance even on minority families.
In contrast, the AST method shows a much wider gap (0.88 weighted vs.\ 0.65 macro), revealing that structural features alone fail to discriminate rare families.
The n-gram method falls between (0.92 weighted vs.\ 0.79 macro), confirming that lexical patterns capture some family-specific content that structural features miss.
This pattern, where the gap between macro and weighted F1 varies systematically across methods, provides diagnostic insight into \emph{what type} of signal each method relies on.

For author attribution, the pattern is different: all three methods show a narrow macro-weighted gap (CodeBERT: 0.76 vs.\ 0.75; n-grams: 0.60 vs.\ 0.57; AST: 0.54 vs.\ 0.51), indicating that performance is relatively uniform across authors of different prolificacy.
This suggests that authorial style, unlike malware content, does not concentrate in a few dominant classes.

\subsection{Implications for Practitioners}
\label{sec:implications}

\paragraph{Metadata removal is insufficient.}
Stripping author tags and organisational names does not prevent attribution.
Structural patterns in conditions, string definitions, and naming conventions carry author-identifying information that a determined adversary with a modest training corpus can exploit.

\paragraph{Repository-level fingerprinting is near-perfect.}
An adversary who observes a shared rule can, with high confidence, determine which major repository it originates from.
For organisations that contribute exclusively to a single repository, this effectively de-anonymises the contributor.

\paragraph{Temporal analysis amplifies risk.}
Recent rules from standardised repositories are \emph{more} attributable than historical ones.
The trend toward rule quality standardisation, while beneficial for detection efficacy, inadvertently increases OPSEC exposure.

\paragraph{Countermeasures must be multi-layered.}
Both surface-level (n-gram) and structural (AST) methods achieve
above-chance attribution, so no single defence suffices:
auto-formatting defeats n-grams but leaves structural signals
intact, while restructuring conditions defeats AST features but
preserves lexical habits. Effective approaches must address both
layers, candidates include style transfer to a canonical form,
federated detection (sharing models rather than rules), and
differential privacy applied to rule structure. The
Webshell~$\times$~YaraHQ result (33.3\%) shows that strict
organisational templates can provide effective protection, at the
cost of suppressing individual expertise.

\subsection{Limitations}
\label{sec:limitations}

Several limitations should be noted.

First, our dataset covers only three public repositories; private organisational rule sets may exhibit different characteristics.

Second, the per-family experiments are limited to the seven families with sufficient multi-author representation; fifteen of 22 families were skipped due to insufficient data.
Third, we do not evaluate adversarial robustness: a motivated author could deliberately alter their style to evade attribution.

Fourth, our CodeBERT models use single-task training; multi-task or few-shot approaches may yield different results.

Fifth, the AST features are approximated via regular expressions rather than a formal YARA parser; a full AST extraction might yield stronger structural signals.

Sixth, malware family labels were assigned via LLM-assisted categorization rather than ground-truth malware analysis; labeling errors could introduce noise into the Task~4 malware family classification results.
However, this does not affect the primary per-family attribution claim (Tier~1 and Tier~2 experiments), which conditions on family membership and is therefore invariant to the absolute correctness of the family taxonomy.


\section{Conclusion}
\label{sec:conclusion}

YARA rules are shared under the working assumption that stripping
metadata is enough to protect the identity of the organisation that
wrote them. To the best of our knowledge, this paper is the \emph{first}
to test that assumption empirically on YARA rules, at scale, and across
four complementary fingerprint dimensions. While prior source-code
authorship attribution work has focused on general-purpose programming
languages drawn from academic contests or large open-source projects,
no prior study has systematically examined whether cybersecurity
detection rules, YARA, Sigma, or Snort, carry exploitable fingerprints
as \emph{authored artefacts} distinct from the malware they detect.
Our answer is unambiguous: the metadata-stripping assumption is wrong.
Repository origin is almost perfectly recoverable (99\% accuracy),
individual authors can be re-identified well above chance (76\%), and
malware-family specialisation is nearly deterministic (95\%). None of
these signals depend on explicit identifiers; they live in the
\texttt{strings} and \texttt{condition} blocks that any useful rule
must keep.

More importantly, the signal does not disappear under the controls that
a reasonable defender would expect to work. Holding malware content
constant still leaves five of seven families attributable; holding both
content \emph{and} repository constant still leaves four of eight
configurations attributable. The attribution surface is multi-layered,
and no single sanitisation step cuts through it.

The most consequential result is not a number but a chain. By sweeping
the per-author rule cap, we showed that the classifier systematically
merges \texttt{yarGen}-generated rules into Tobias Michalski's class at
the 85--88\% level, and that this ``error'' corresponds to a real, publicly
verifiable institutional coupling: Michalski and Florian Roth are both
affiliated with Nextron Systems, the company that also maintains
\texttt{yarGen} and the Neo23x0 repository. What looks like a confusion
between a tool and an analyst is in fact the classifier recovering the
\emph{institution} that owns both. This is the attribution threat that
ISAC and NIS2-style sharing programs have not yet confronted:
institutional fingerprints aggregate signal from the tool, the
analysts, the linters, and the review process simultaneously, and they
cannot be removed by individual anonymisation.

We frame our temporal result, a 9--20 percentage point accuracy gap between the full
history and the 2022--2025 subset, as a premise rather than a
conclusion. The effect is consistent in direction and magnitude across
three independent methods, but it rests on a single repository's
commit history and a four-year window. We therefore offer it as
suggestive evidence that fingerprint strength is not static over time,
and we call for longer-horizon studies on corpora with richer temporal
annotation to test it.

The operational implication for the YARA-sharing ecosystem, and for the
sector-specific ISACs and national CERTs that anchor it, is that
contributor anonymity cannot be treated as a property of the rule
header. It is a property of the entire production pipeline that
produced the rule, and defending it requires intervention at every
stage of that pipeline: the choice of generator, the structure of
templates, the enforcement of style guides, and ultimately the
willingness to share representations of rules rather than rules
themselves. Our findings motivate a research programme on
privacy-preserving detection-rule sharing in which attribution
resistance is treated as a first-class design objective, alongside
detection coverage and operational efficiency. The dataset, the
experimental pipeline, and the attribution-verdict framework introduced
here are published to support that programme.




\bibliographystyle{ACM-Reference-Format}
\bibliography{references}

\appendix
\section{Open Science}
\label{sec:open-science}

In line with the CCS~2026 Open Science policy, all artefacts required
to reproduce the results reported in this paper are released at an
anonymous URL for the duration of the review process. The underlying
YARA rules are drawn from three publicly available GitHub repositories
(\texttt{Neo23x0/signature-base}, \texttt{reversinglabs-yara-rules},
and \texttt{YARAHQ/yara-forge}), whose upstream licences are preserved
in the released bundle.

\paragraph{Anonymous artefact URL.}
\url{https://anonymous.4open.science/r/anon-yara-attribution-XXXX}%
\footnote{Placeholder; the final anonymous URL is provided in the
HotCRP submission form and will be active from the submission date
through the end of the review cycle.}

\paragraph{Released artefacts.} The repository contains:
\begin{itemize}
  \item {Corpus and preprocessing.} The unified dataset
  yara4\_with\_repo.csv ($23{,}305$ rules; columns
  {text}, {author}, {malware\_family},
  {repo}, {rule\_name}) and the time-restricted subset
  {yara2.csv} ($14{,}366$ rules from 2022--2025,
  Neo23x0/signature-base) used for author attribution and timed
  repository attribution. The preprocessing pipeline implementing
  metadata stripping, block-comment removal, tag removal, and
  class-size filtering as described in~\S5.2 is provided as a
  reproducible script.

  \item {Malware-family label assignment.} The two-step
  LLM-assisted categorisation pipeline used to label
  Malware family~(\S5.1), including the fixed
  taxonomy of 22 families, and the produced rule-to-family mapping.

  \item {Classifier implementations.}
  (i)~the shared CodeBERT classifier module
  ({codebert\_common.py}), implementing the
  {CodeBERTClassifier} and {SingleTaskDataset} used
  across all four tasks;
  (ii)~four single-task training drivers
  ({task1\_author.py}, {task2\_repo\_timed.py},
  {task3\_repo\_full.py}, {task4\_malware.py});
  (iii)~lexical $n$-gram and AST-feature baselines \\
  (ast\_ngrams\_training.py) reproducing the Burrows'
  Delta and Caliskan-Islam-style pipelines reported in Table~1.

  \item {Confound-controlled experiments.} The Tier~1
  per-family and Tier~2 per-family\,$\times$\,per-repository
  training scripts (Tables~2--3), the varying-$N$ sweep script
  used for the Hacktool analysis in~\S6.3, and the
  confusion-matrix extraction code used for Figure~8.

  \item {Figure and table regeneration.} Scripts that
  regenerate every table and figure in the paper
  (Tables~1--4; Figures~2--8) from the logged training outputs,
  with data and plotting parameters separated from plotting
  logic for ease of inspection.

  \item {Training logs.} Per-task stdout/stderr logs
  recording hyperparameters, per-epoch metrics, early-stopping
  events, and validation scores, supporting independent
  verification of the reported numbers without re-running
  training.

  \item {Reproduction instructions.} A {README}
  specifying the conda environment
  ({PyTorch~2.6.0+cu118}, CUDA~11.8, Transformers,
  scikit-learn), the command order, the GPU-adaptive execution
  mode with a CPU fallback, and the expected runtime per task
  on a single GPU.
\end{itemize}

\paragraph{Model checkpoints.} We do not release fine-tuned
CodeBERT checkpoints as part of the anonymous artefact bundle.
Training is fully deterministic given the released code, the
fixed corpus (yara4\_with\_repo.csv), and the
hyperparameters specified in \S5.3 (AdamW, learning rate
$2\times10^{-5}$, batch size 16, early stopping with patience~3,
seed fixed in the configuration file). Every reported accuracy,
macro-F1, and weighted-F1 value in Tables~1--4 can be
regenerated end-to-end from the released training drivers.
Per-task training logs recording per-epoch metrics,
early-stopping events, and final validation scores are included
in the bundle for independent verification without re-running
training.

\paragraph{No human-subjects data.} The corpus contains only the
source text of YARA detection rules published in public GitHub
repositories. No personal data, no private repositories, and no
endpoint or sandbox telemetry are used. Author and repository
labels originate from the public commit history of the respective
repositories and are not the product of any deanonymisation
performed by the authors.

\section{Ethical Considerations}
\label{sec:ethics}

This paper studies a privacy risk inherent to the sharing of YARA
detection rules. We consider the ethical implications below,
following the USENIX Security'26 ethics guidance referenced by
the CCS~2026 CFP.

\paragraph{Data sources and consent.} All rules used in this study
are drawn from three public GitHub repositories, each operating
under a permissive open-source licence that explicitly allows
redistribution and analysis. Author and repository labels are not
inferred or reconstructed: they are already present, in plaintext,
in the public commit history and in the \texttt{meta:} fields of
the unredacted rules as published by their authors. No private
repositories, private feeds, ISAC-restricted material, or
TLP-restricted advisories were collected or analysed. Consequently,
no individual or organisation is identified in this work beyond
what they have already chosen to publish under their own name.

\paragraph{Identification of named individuals and organisations.}
Our institutional-attribution finding~(\S6.3) discusses two
analysts (Tobias Michalski and Florian Roth) and one organisation
(Nextron Systems) by name. Each of these names already appears as
an \texttt{author} field in the public rules of
\texttt{Neo23x0/signature-base} and in publicly authored technical
posts on the organisation's official blog, both of which we cite.
We name them only to make the institutional-coupling argument
externally verifiable by the reader; no additional private
information is introduced, and no claim is made about either
individual beyond their publicly documented affiliation with
Nextron Systems and their role in maintaining the
\texttt{Neo23x0/signature-base} repository and the \texttt{yarGen}
tool. We do not attribute any specific rule to either analyst
beyond the attribution already present in the upstream repository.

\paragraph{Risk--benefit analysis.} The central contribution of
this work is a demonstration that metadata removal alone does not
preserve contributor anonymity in YARA rule sharing. We considered
whether releasing this result could expose specific contributors
or organisations to new risk, and concluded that the net effect
is defensive.

\begin{itemize}
  \item The attribution capability demonstrated here requires only
  publicly scrapeable rule corpora and standard ML tooling. Any
  adversary motivated to re-identify rule authors already has
  access to the same inputs and the same methods. Our paper does
  not introduce a new capability; it quantifies one that already
  exists and is unlikely to have gone unexploited.

  \item Defenders, ISACs, national CERTs, and the NIS2-aligned
  operators of essential services who coordinate detection
  sharing, currently treat metadata stripping as sufficient
  protection. Withholding this result would leave that assumption
  unchallenged. Publishing it allows the community to design
  countermeasures~(\S7.5) and to set realistic expectations for
  contributor anonymity in future sharing frameworks.

  \item The specific institutional-attribution example exposes a
  coupling between an openly maintained rule generator
  (\texttt{yarGen}), an openly maintained repository
  (\texttt{Neo23x0/signature-base}), and an openly branded
  organisation (Nextron Systems). This coupling is documented on
  the organisation's own public channels. We do not believe its
  discussion creates any reputational, legal, or operational risk
  beyond what is already inherent in the organisation's chosen
  public profile.
\end{itemize}

\paragraph{Notification.} Our findings concern information that
Nextron Systems, Mr.~Michalski, and Mr.~Roth have themselves chosen
to publish, on the organisation's official blog and in the authored
rules of the \texttt{Neo23x0/signature-base} repository. The
institutional coupling we describe is documented on Nextron Systems'
own public channels and is cited accordingly. We have therefore not
treated this work as a vulnerability disclosure, and no embargo or
pre-publication notification procedure applies. At the same time, we
recognise that naming individuals in a re-identification paper
warrants caution. Accordingly, (i)~no claim in the paper goes beyond
what is verifiable from the cited public sources, (ii)~no specific
rule is attributed to an individual beyond the attribution already
present in the upstream repository, and (iii)~the paper's framing is
squarely defensive, quantifying an attribution surface that ISAC and
NIS2-aligned sharing programmes should design against rather than
providing a novel offensive capability.

\paragraph{Dual-use considerations.} The released artefacts
(corpus, training code, labels) allow reproduction of our
attribution classifiers. We assessed whether release itself poses
incremental offensive risk and concluded it does not: the
classifiers operate on public data that any adversary can already
collect, and the primary audience for the released code is the
defender community developing countermeasures. The released
pipeline can equally be used to audit the attribution resistance
of a candidate rule-sharing protocol, which is an intended use.

\paragraph{LLM-assisted data labelling.} Malware-family labels
for the \texttt{yara3} subset were produced by a frontier LLM
applied to the \texttt{rule\_name} and \texttt{meta:
description} fields of each rule~(\S5.1). These labels are used
only as coarse-grained stratification for the per-family
experiments; the core attribution claims (Tiers~1 and~2) are
invariant to the absolute correctness of the family taxonomy, as
they condition on family membership rather than predicting it.
We flag this choice explicitly so that reviewers and downstream
users can weight the Task~4 malware-classification numbers
accordingly.

\paragraph{No IRB/ERB review.} The study does not involve human
subjects, user data, or interaction with third parties. IRB/ERB
review is therefore neither required nor meaningful under our
institution's definition of human-subjects research. We have
nonetheless reasoned about the ethics of the work beyond formal
institutional compliance, as reflected in this section.
\end{document}